%% file: ASBH.tex
\documentclass[aps,prd,onecolumn,groupedaddress,showpacs,nofootinbib,amssymb,notitlepage]{revtex4-1}
\pdfoutput=1
\usepackage[usenames,dvipsnames]{color}
\usepackage[colorlinks=true, linkcolor=Black, citecolor=Black, urlcolor=Black, filecolor=Black]{hyperref}
\usepackage{longtable}
\usepackage{soul}
\usepackage{epsfig}
\usepackage{amssymb}
\usepackage{amsmath}
\usepackage{graphicx}
\usepackage{verbatim}
\usepackage{xspace}
\usepackage{hyperref}
\input{jdefs}
\newcommand{\nc}{\newcommand}
\nc{\be}{\begin{equation}}
\nc{\ee}{\end{equation}}
\nc{\ba}{\begin{eqnarray}}
\nc{\ea}{\end{eqnarray}}

\def\bc{\begin{center}}
\def\ec{\end{center}}

\def\to{\rightarrow}

\def\bc{\begin{center}}
\def\ec{\end{center}}

\begin{document}

\title{Towards conditions for black-hole singularity-resolution \\in asymptotically safe quantum gravity}
\author{Ad\'em\d{\'o}l\'a Ad\'e\`if\d{\'e}\d{o}ba}
\email{a.adeifeoba@thphys.uni-heidelberg.de}
\author{Astrid Eichhorn}
\email{a.eichhorn@thphys.uni-heidelberg.de}
\author{Alessia Platania}
\email{a.platania@thphys.uni-heidelberg.de}
\affiliation{Institut f\"ur Theoretische Physik, Universit\"at Heidelberg, Philosophenweg 16, 69120 Heidelberg, Germany}

\begin{abstract}
We explore the fate of the curvature singularity of Schwarzschild (deSitter) black holes in asymptotically safe quantum gravity. Specifically, we upgrade the classical spacetime by including the running of the Newton coupling and cosmological constant. In this setting, the antiscreening character of the gravitational interaction can remove the singularity, yet a nonzero value of the cosmological constant in the ultraviolet appears to reintroduce it. We find hints that a finite value of the cosmological constant in the infrared is compatible with singularity resolution
provided that the cosmological constant is driven to zero fast enough in the ultraviolet. We compare the corresponding bounds on the critical exponents to the literature.
\end{abstract}

\maketitle

\section{Introduction}

 The observation of gravitational waves by LIGO in 2015 \cite{Abbott:2016blz} is compatible with the predictions of General Relativity (GR) for the properties of spacetime. At the same time, it reinforces the need to go beyond GR for a full understanding of spacetime across all scales. The latter conclusion follows by asking for the source of the signal GW1509, which according to GR is a black-hole binary. The curvature singularity inside black holes, which renders the  spacetime geodesically incomplete, highlights the necessity to extend GR at very high curvature scales.
The most widely accepted assumption is that quantum gravitational effects resolve the black-hole singularity and provide a predictive, fundamental description of gravity.
In this paper, we discuss the conditions for singularity-resolution in Schwarzschild-deSitter black holes modified by the running of the gravitational couplings triggered by quantum fluctuations. Our analysis will be based on the assumption that there is a scale-invariant regime that is reached at the Planck scale, realizing the asymptotic safety scenario for quantum gravity proposed by Weinberg \cite{1976W}. Indications for asymptotic safety come from studies of the truncated Renormalization Group (RG) flow of gravity in a Riemannian setting \cite{martin,Souma:1999at,Lauscher:2001rz,martinfrank,Litim:2003vp,Donkin:2012ud,Lauscher:2002sq,Codello:2007bd,Machado:2007ea,Benedetti:2009rx,Manrique:2009uh,Groh:2010ta,Eichhorn:2010tb,Manrique:2011jc,Benedetti:2012dx,Dietz:2012ic,Christiansen:2012rx,Falls:2013bv,Becker:2014qya,Falls:2014tra,Christiansen:2014raa,Gies:2015tca,Demmel:2015oqa,Ohta:2015efa,Falls:2016msz,Gies:2016con,Denz:2016qks,AleFrank1,Knorr:2017fus,Christiansen:2017bsy,deBrito:2018jxt}, and even in the presence of Standard Model matter fields \cite{Dona:2013qba,Dona:2014pla,peva14,Percacci:2015wwa,Meibohm:2015twa,Oda:2015sma,Dona:2015tnf,Labus:2015ska,Eichhorn:2016vvy,Hamada:2017rvn,Eichhorn:2017sok,AleFrank2}, with intriguing hints for a quantum-gravity induced UV-completion of the Standard Model with an enhanced predictive power \cite{Shaposhnikov:2009pv,Daum:2009dn,Harst:2011zx,Folkerts:2011jz,Eichhorn:2011pc,Eichhorn:2016esv, Meibohm:2016mkp,Eichhorn:2017ylw,Eichhorn:2017lry,Eichhorn:2017als,Christiansen:2017cxa,Eichhorn:2018whv}.

Modifications of black-hole spacetimes induced by quantum gravity could also go beyond singularity resolution, and even generate potentially observable consequences. Whether this is indeed the case, depends on the form of the effective action for quantum gravity. It encodes the effects of quantum fluctuations at all scales and is expected to contain additional contributions beyond the Einstein-Hilbert term. Therefore, constraints on  gravity arising from gravitational-wave observations also constrain those modifications with a quantum origin, and thereby, quantum-gravity models.
At leading order in the curvature expansion, the additional terms are those explored in \cite{Stelle:1977ry}. The vacuum solutions of the corresponding equations of motion can still contain the black-hole spacetimes of GR, but can also include additional solutions \cite{Lu:2015cqa}. Moreover, the equations describing linear perturbations of the spacetime, i.e., the quasi-normal modes of black holes, could differ from those derived from the Einstein equations \cite{Berti:2018vdi}.  Along these lines, a confrontation of asymptotically safe quantum gravity with observational data might become possible in the future. A first step in this program is to understand the static solutions, and whether these are singularity-free, before moving on to derive quasi-normal-mode spectra and further quantities potentially closer to observations than the spacetime structure behind the horizon. The main aim of our work will be reconciling singularity-resolution in asymptotically safe quantum gravity with the existence of a cosmological constant, which has proven to be a major challenge \cite{ko14, Pawlowski:2018swz}.

In order to explore the question of singularity-resolution in asymptotically safe quantum gravity, we neglect several ingredients key to realistic astrophysical systems, such as angular momentum of the black hole and the presence of matter and radiation. Furthermore, as the causal structure of the spacetime plays a central role in black- hole physics, we  work in a Lorentzian setting, and make the assumption that Lorentzian quantum gravity does indeed become asymptotically safe, see \cite{Manrique:2011jc}. We follow an RG improvement procedure to upgrade the Schwarzschild-deSitter solution to a metric that is expected to capture leading-order quantum effects. Hence the latter is no longer a solution to the vacuum Einstein equations with cosmological constant. The effective energy-momentum tensor that is required to make the metric a solution to the Einstein equations can be interpreted as encoding quantum gravitational effects. While the RG improvement is not a rigorous derivation of quantum gravity effects, it is expected to capture qualitative features of the corresponding quantum-corrected spacetime \cite{cw,1973migdal,1973gross,1978pagels,1978Matinyan,1983adler,br02,2004reuterw1,2004reuterw2,alfio12,AlfioAle1}, especially when the system is characterized by a high degree of symmetry. In particular, we limit ourselves to an analysis of the structure of spacetime at very high curvatures, close to the classical singularity, where the gravitational RG flow approaches a scale-invariant regime and the behavior of gravity is controlled by universal quantities only, such as critical exponents.

Due to the intrinsic interest of the fate of the curvature singularity of black holes and the feasibility of studies along the lines outlined above, the literature on ``RG-improved'' black holes in asymptotic safety is rather extensive \cite{1999alfiomartin,br00,2006alfiomartin,Falls:2010he,fayos11,Falls:2012nd,torres14%
,torres14b,ko14,BKP,torres15,Kofinas:2015sna,Bonanno:2017zen,Pawlowski:2018swz}. It has resulted in ``quantum-corrected'' black holes including those characterized by a regular metric in their interior. Yet, a key question has emerged when upgrading studies based on the running Newton coupling by a running cosmological constant. As shown in \cite{ko14}, this upgrade brings back the singularity. This is the main motivation for our work, in which we explore the conditions that arise on the fixed-point properties in asymptotic safety, if a scale-dependent cosmological constant is included and singularity resolution is required.

This paper is organized as follows. In Sect.~\ref{sec:QIBH} we review the basic formalism.
In Sect.~\ref{sec:linFP} we recapitulate the scaling behavior of the running couplings in the ultraviolet (UV) regime, where the RG flow of gravity could be controlled by an interacting fixed point.
The impact of the critical exponents on the singularity-structure of  Schwarzschild-deSitter black holes is then presented
in detail in Sect.~\ref{sect3}. Finally, in Sect.~\ref{sect4} we summarize our findings and discuss future extensions.

\section{RG improved black holes} \label{sec:QIBH}

Let us consider the following spherically symmetric spacetime {in Schwarzschild coordinates}
\be\label{Cmetric}
ds^2=-f(r)dt^2+\frac{1}{f(r)}dr^2+r^2d\Omega^2\;\;.
\ee
Here $\Omega$ is the solid angle in four spacetime dimensions and the lapse function $f(r)$ is given by
\be\label{eq:classlapse}
f(r)=1-\frac{2\,m\,G_0}{r}-\frac{\Lambda_0}{3} r^2\;\;,
\ee
where $G_0$ and $\Lambda_0$ are the classical Newton 
coupling and cosmological constant, respectively. The Kretschmann scalar constructed using the metric \eqref{Cmetric} reads
\be\label{eq:ClassKre}
K_0= R_{\mu\nu\kappa\lambda}R^{\mu\nu\kappa\lambda}=\frac{48 \, m \, G_0}{r^6}+{\frac{8\Lambda_0^2}{3}}\;\;,
\ee
and diverges as $r\to0$. The metric \eqref{Cmetric} is thus characterized by a curvature singularity and the resulting spacetime is geodesically incomplete. 

We assume that the impact of leading-order quantum gravitational effects on
the structure of black holes can be studied using a Renormalization Group improvement procedure \cite{br00}. The latter is a standard technique employed in Quantum Field Theory \cite{cw,1978pagels,1978Matinyan} and consists of replacing the coupling constants of the systems with the corresponding running couplings
\be
G_0 \rightarrow G(k),\quad \Lambda_0 \rightarrow \Lambda(k)\;\;,
\ee
where $k$ is the RG scale, and subsequently identifying $k$ with a physical scale of the system. This method has also provided insight into the spectral dimension of asymptotically safe spacetimes, indicating dynamical dimensional reduction \cite{Lauscher:2005qz,Reuter:2011ah,Rechenberger:2012pm,Calcagni:2013vsa}. Similarly, graviton-mediated scattering cross-sections have been estimated in this way \cite{Litim:2007iu,Gerwick:2011jw,Dobrich:2012nv}. In settings such as the present one, the high degree of symmetry of the spacetime results in the presence of a unique high-energy scale, namely the curvature scale. This provides an essentially unique way of performing the RG improvement in the ultraviolet regime, different from cases with several distinct physical high-energy scales. 

The first step of the procedure results in a $k$-dependent metric of the form \eqref{Cmetric}, with an RG-improved lapse function given by
\be \label{rglapse}
f_k(r)=1-\frac{2m\,G(k)}{r}-\frac{\Lambda(k)}{3} r^2\;\;,
\ee
where the scale dependence of $G(k)$ and $\Lambda(k)$ is determined by solving the corresponding beta functions with suitable initial conditions.

The second step of the procedure consists in performing a scale identification of the form
\be\label{eq:identification}
k = k[\mathcal{S}_{\rm phys}(r)]\;\;,
\ee
where $\mathcal{S}_{\rm phys}(r)$ is a physical scale of the system,  and the specific functional form of $k[\cdot]$ is dictated by dimensional arguments. We note that the cosmological constant is an IR scale, and therefore not expected to be relevant for a physically meaningful scale-setting aimed at the analysis of the singularity structure. In other words, the identification between the RG scale $k$ and a physical high-energy scale of Schwarzschild-deSitter spacetime should not depend on $\Lambda$. As advertised, seemingly different choices for the identification, e.g., $k = K_0^{\alpha_1}$ or $k=d_0(r)^{\alpha_2}$, where $d_0(r)$ is the distance along a radial geodesic and $\alpha_{i}$ are determined by dimensional arguments, lead to the same result for the quantum-``improved'' spacetime, cf.~Sect.~\ref{sec:fullscaleid}.

\section{Linearized flow of gravity} \label{sec:linFP}

In this section we summarize the properties of the gravitational RG flow in the trans-Planckian regime, $k\gtrsim M_\mathrm{P}$. Note that $M_{\rm P}$ is to be understood as a placeholder for the transition scale to the fixed-point regime, which appears to approximately agree with the Planck scale in simple approximations \cite{Reuter:2004nx}, but a priori could also be lower/higher.
In this regime the gravitational RG flow is controlled by the 
interacting fixed point. In the approach to it, the scaling of 
the (dimensionless) couplings $\mathbf{g}\equiv \{g^1,\dots,g^n\}$  is completely determined by the corresponding critical exponents. The couplings are made dimensionless through an appropriate rescaling with $k$ and the scaling of their dimensionful counterparts follows straightforwardly. Expanding the beta functions $\beta^i$ around a fixed point $\mathbf{g}_\ast\equiv \{g^1_\ast,\dots,g^n_\ast\}$ of the RG flow yields
\be\label{flow:lin}
k \partial_k g^i (k) = \sum_{j=1}^{n} \left.\frac{\partial \beta^i}{\partial g^j}\right|_{\mathbf{g}_\ast} \hspace{-0.1cm} (g^j(k) - g_\ast^j) +\mathcal{O}\left(g^j(k)-g^j_{\ast}\right)^2.
\ee
To linear order, the solution reads
\be\label{eq:linsol}
g^i(k) = g_\ast^i + \sum_{j=1}^n c_j \, 
V_j^i \left( \frac{k}{M_\mathrm{P}} \right)^{-\theta_j}\;\;.
 \ee
Here $V^i_j$ denotes the $i$th component of the $j$th eigenvector, $\theta_i$ are the critical exponents and the $c_i$ are integration constants.  The linearization captures both infrared (IR) repulsive as well as IR-attractive directions of the fixed point, with $\mathrm{Re}[\theta_j]>0$ being the condition for a IR-repulsive (i.e., UV-attractive) direction. These are the relevant directions. For a trajectory that emanates from the fixed point in the UV, all $c_i$ corresponding to $\mathrm{Re}[\theta_i]<0$ are set to zero, as those directions are not UV-attractive. This renders a fixed point a predictive UV completion if the number of relevant directions is finite.
The linearized solution~\eqref{eq:linsol} describes the scaling of couplings in the fixed-point regime and breaks down elsewhere.

Here we focus on a truncation of the gravitational RG flow following a canonical ordering up to dimension-2-operators with a running Newton coupling and cosmological constant.
The Einstein-Hilbert truncation is not expected to capture the full dynamics of asymptotically safe quantum gravity, as higher-order curvature terms are generated by quantum corrections. According to many studies, an additional coupling at the curvature-squared level is relevant in a setting with only gravitational degrees of freedom \cite{Lauscher:2002sq,Machado:2007ea,Benedetti:2009rx,Falls:2013bv,Falls:2014tra,Denz:2016qks,Christiansen:2017bsy,deBrito:2018jxt} and thereby contributes an additional free parameter to the Lagrangian. In a setting where quantum fluctuations of matter are additionally taken into account and contribute to the critical exponents for gravity, the number of relevant directions is not settled yet, but might even be lower \cite{Alkofer:2018fxj}. On the other hand, the Schwarzschild spacetime remains a solution in much more general settings, where $R_{\mu\nu}=0$. For instance, this is the case for quadratic gravity as well as $f(R)$ gravity. In addition, Schwarzschild-deSitter black holes are solutions to the field equations of quadratic-gravity theories. Therefore, although the couplings of higher-order terms in the Lagrangian could modify the beta functions of 
$G_k$ and $\Lambda_k$, they do not necessarily impact the structure of the RG-improved solution~\eqref{rglapse} (cf.~Sect.~\ref{sec:realtheta}). The study of the RG improvement of additional, non-Schwarzschild solutions of higher-derivative gravity \cite{Lu:2015cqa} will not be subject of this paper.

For our purposes it is therefore sufficient to consider the 
Einstein-Hilbert subspace, spanned by the dimensionless Newton coupling 
\be
g(k)\equiv G_k \,k^{2},
\ee
 and cosmological constant 
 \be
 \lambda(k)\equiv \Lambda_k\,k^{-2}.
 \ee 
 In particular, we discuss the case of real and complex  critical exponents separately. In the former case, the running cosmological constant and Newton coupling read
\begin{subequations}  \label{scalingrc}
\begin{align} 
\lambda(k) = \lambda_\ast &+ \lambda_1 \, \left(\frac{k}{M_\mathrm{P}}\right)^{-\theta_1}+\lambda_2 \, \left(\frac{k}{M_\mathrm{P}}\right)^{-\theta_2} \;\;, \\
g(k) =g_\ast &+g_1 \, \left(\frac{k}{M_\mathrm{P}}\right)^{-\theta_1}+g_2 \, \left(\frac{k}{M_\mathrm{P}}\right)^{-\theta_2} \;\;. \label{ggsc}
\end{align}
\end{subequations}
Herein, $\lambda_i$ and $g_i$ are free parameters, cf.~Eq.~\eqref{eq:linsol}.
In the case of a complex pair of critical exponents, $\theta_\pm=\theta'\pm i \theta''$, the two couplings can be 
expressed in the following form 
\begin{subequations}  \label{scalingrc2}
\begin{align}
\lambda(k) = & \,\,\lambda_\ast  + \, \big(\lambda_1 \cos(\theta^{\prime\prime} t) + \lambda_2 \sin(\theta^{\prime\prime} t)   \big) 
\left( \frac{k}{M_\mathrm{P}} \right)^{-\theta^\prime} \;\;, \\ \label{ggsc2}
g(k) = & \,\,g_\ast  + \, \big(g_1 \cos(\theta^{\prime\prime} t) + g_2 \sin(\theta^{\prime\prime} t)   \big)  
\left( \frac{k}{M_\mathrm{P}} \right)^{-\theta^\prime} \;\;,
\end{align}
\end{subequations}
where $t \equiv \ln(k/M_\mathrm{P})$. In the context of asymptotically safe quantum gravity both these cases could be of interest since, depending on the choice of truncation and regulator, both cases occur in the literature (see~Tab.~\ref{tab:FPresults}).

\section{The role of critical exponents for singularity resolution} \label{sect3}

In our setup the quantum corrections to the classical Schwarzschild-deSitter spacetime are
encoded in the scale-dependent lapse function \eqref{rglapse}. The latter can be written in terms  of the dimensionless counterpart of the running Newton coupling and cosmological constant
\be \label{rglapse2}
f_k(r)=1-\frac{2\,m\,g(k) \,k^{-2}}{r}-\frac{\lambda(k)\,k^2}{3} r^2\;\;.
\ee
In the regime $k\gtrsim M_\mathrm{P}$, corresponding to the region surrounding the classical singularity, the functions $\lambda(k)$ and $g(k)$ are determined by the linearized beta functions discussed in the previous section. 
The complete RG-improved metric is obtained by specifying the functions $g(k)$ and $\lambda(k)$, as well as the cutoff function $k(r)$, and inserting them into Eq.~\eqref{rglapse2}. 
Based on the intuition that quantum-gravity effects are irrelevant at large distance scales and set in close to the classical singularity, the RG-scale $k(r)$ must be a monotonically decreasing function of the radial coordinate $r$. In particular, as we are interested in the spacetime structure in the regime $k\gtrsim M_\mathrm{P}$, corresponding to $r\ll1$, the RG cutoff $k(r)$ can be written as
\be \label{scalk}
{k(r)}\simeq \tilde{\xi}\,M_\mathrm{P} \,r^{-\gamma} \;\;.
\ee
Here $\gamma$ is a positive number and $\tilde{\xi}$ is an appropriate dimensionful parameter given by
\be
\tilde{\xi}=\frac{\xi}{M_{\rm P}} \,(m\,G_0)^{\gamma-1}\;\;,
\ee
where  $\xi$ is an arbitrary, positive constant. Note that for $\gamma=3/2$ the cutoff identification \eqref{scalk} reduces to the physical scale $k(r)=\xi\, d^{-1}_0(r)$ \cite{br00,2006alfiomartin}, where $d_0(r)$ is the proper  distance obtained from the expansion of the classical Schwarzschild-deSitter metric about $r=0$. We highlight that other choices of $\gamma$ do not admit a coordinate-invariant interpretation.

In our setting, the running couplings fundamentally modify the spacetime structure close to the  classical singularity  and induce a modified lapse function
\be\label{modNew}
f(r) =1-2\phi(r)=1-2\phi_G(r)-2\phi_\Lambda(r)\;\;,
\ee
with
\be
\phi_G(r)=\frac{m\,G(r)}{r}\,,\qquad \phi_\Lambda(r)=\frac{\Lambda(r)}{6}r^2\;\;.
\ee
The notation already suggests that in the weak-field regime far away from the horizon, the $r$-dependent part of the lapse function reduces to the Newtonian potential.
The modified Ricci and Kretschmann scalars read
\begin{subequations} \label{curvature}
\begin{align} 
&R=\frac{4\phi}{r^2}+\frac{8\phi'}{r}+2\phi''\;\;,\\
&K=\frac{16\phi^2}{r^4}+\frac{16\phi'^2}{r^2}+4\phi''^2\;\;.
\end{align}
\end{subequations}
For $G(r)= G_0$ and $\Lambda(r)= \Lambda_0$, the Ricci scalar reduces to $R= 4 \Lambda_0$, as expected. The RG flow of the running gravitational couplings allows to remove the classical singularity when the $r$-dependent part of the modified  lapse function in Eq.~\eqref{modNew} is such that $\phi^{(n)}(r)$ vanishes no slower than $r^{n-2}$ as $r\to0$. The latter condition is satisfied when the leading order scaling of $\phi(r)$ is
\be\label{conditionreso}
\phi(r)\simeq r^\delta\;\;, \qquad \delta\geq2\,,
\ee
for small $r$. 

We now investigate the constraints for the critical exponents and derive a bound on $\gamma$ such that the condition \eqref{conditionreso} for singularity-resolution is fulfilled. 
\subsection{Case of real critical exponents}\label{sec:realtheta}

In the case of real critical exponents, the RG-improved metric is obtained by combining the scaling of the dimensionless gravitational couplings in Eq.~\eqref{scalingrc} with the running lapse function in Eq.~\eqref{rglapse2}. The scaling of the RG-improved lapse function~\eqref{modNew} about $r=0$ thereby depends on the critical exponents $\theta_{1,2}$. In particular, the modified $\phi_G$-function reads
\be
\phi_G(r)\equiv \frac{m}{\tilde{\xi}^{2}M_{\rm P}^2} \left(g_\ast+(g_1 \tilde{\xi}^{-\theta_1})\,r^{\gamma\,\theta_1}+(g_2 \tilde{\xi}^{-\theta_2})\,r^{\gamma\,\theta_2}\right) \, r^{2\gamma-1}\;\;.
\ee
According to \eqref{conditionreso}, $\phi_G(r)$ gives a non-singular contribution to the curvature invariants \eqref{curvature} when $\gamma\geq3/2$. We  discuss physically motivated scale-identifications that result in $\gamma \geq 3/2$ in Sect.~\ref{sec:fullscaleid}. 
A first restriction on the critical exponents arises, namely  $\theta_{1,2}> \frac{3-2\gamma}{\gamma}$, i.e., $\theta_{1,2}\geq 0$ for $\gamma=3/2$. At a first glance, one might not expect the critical exponents to play a role in the discussion of the singularity, as the far UV regime is expected to correspond to the fixed-point regime, whereas the critical exponents characterize the approach to the latter. Yet for $\theta_i<0$ and $g_{1,2}\neq0$, the corresponding RG trajectory is not UV safe, and the Newton coupling diverges at high energies. This becomes clear from Eq.~\eqref{eq:linsol}: $\theta_i<0$ corresponds to a UV repulsive direction, such that an independent choice of the low-energy value of the coupling is not possible for a UV safe trajectory.
This divergence  results in the occurrence of a new singularity due to a UV-divergent $G$. On the other hand, any $\theta_i>0$ is associated to a UV safe Newton coupling $g(k)\rightarrow g_{\ast}$. This implies a weakening of the dimensionful gravitational interaction strength through antiscreening quantum-gravity fluctuations, i.e., $G(k) = g_{\ast}/k^2$. It seems likely that this should result in a weakening or resolution of the classical singularity.

The modifications to $\phi_{\Lambda}$ are instead encoded in
\be
\phi_\Lambda(r)\equiv \frac{\tilde{\xi}^{2}M_{\rm P}^2}{6} \left(\lambda_\ast+(\lambda_1 \tilde{\xi}^{-\theta_1})\,r^{\gamma\,\theta_1}+(\lambda_2 \tilde{\xi}^{-\theta_2})\,r^{\gamma\,\theta_2}\right) \, r^{2-2\gamma}\;\;.
\ee
As $\gamma$ is positive, the resolution of the singularity requires the cosmological constant to vanish in the ultraviolet limit, $\lambda_\ast=0$. The existence of a singularity for $\lambda_{\ast}\neq 0$ has already been discussed in \cite{ko14}: if only the fixed-point scaling of $g (k)$ and $\lambda (k)$ is taken into account and the scale-identification \eqref{scalk} with $\gamma=3/2$ is used, then the $\phi_G$ and $\phi_\Lambda$ terms in Eq.~\eqref{modNew} generate the following lapse function
\be \label{treelapse}
f_\ast(r)=1-\frac{2\, g_\ast M_\mathrm{P}^2}{\xi^2 } \,r^2-\frac{m\, \xi^2\lambda_\ast }{3rM_\mathrm{P}^2}\;\;.
\ee
Herein, the Newton coupling and cosmological constant reverse their roles in the region surrounding the classical singularity. In other words, when the RG flow reaches the fixed-point regime, the cosmological constant turns into an effective Newton coupling and, in turn, the running Newton coupling ``back-reacts'' by generating a deSitter core \cite{Falls:2010he,ko14}. Accordingly, $\lambda_{\ast}\neq 0$ would lead to a Kretschmann scalar diverging as in the classical case, i.e., as $r^{-6}$. Thus a condition for singularity resolution in this setting is $\lambda_{\ast}=0$.

However, our universe is not asymptotically flat and a non-zero cosmological constant exists. How can this be reconciled with the requirement $\lambda_{\ast}=0$? The answer is actually simple: in our setup, is it only the UV-value of the cosmological constant which has an impact on the microscopic structure of spacetime. Yet, the observations requiring a non-vanishing cosmological constant pertain to its IR value. Hence, a \emph{running} cosmological constant, which vanishes in the UV, but is generated by quantum fluctuations along the RG flow towards the IR regime can meet both constraints.

As a cosmological constant has to emerge dynamically when the RG-scale $k$ is lowered towards the infrared regime (asymptotic region, $r\gg 1/M_{\rm Planck}$), in this paper we focus not only on the fixed-point regime but also on the linearized approach to it. Setting $\lambda_\ast=0$ and applying the condition~\eqref{conditionreso} to $\phi_\Lambda(r)$, one sees
that a non-singular Schwarzschild-deSitter spacetime characterized by an \textit{emergent cosmological constant} can only be realized when $\gamma(\theta_i-2)\geq0$. Remarkably, the latter condition does not depend on the particular scaling relation~\eqref{scalk} employed. Given that $\gamma$ must be positive, it implies the following constraint
\be\label{eq:condtheta}
\theta_i\geq2\;\;.
\ee
Again, one might wonder why the existence of a singularity at $r= 0$ is sensitive to the approach to the fixed point, rather than the fixed-point behavior only. For the cosmological constant, singularity resolution requires it to be asymptotically free, which means that it is actually finite at any finite scale $k$. Unless it approaches zero fast enough (encoded in the condition $\theta\geq2$), a new, quantum-gravity induced singularity is generated. 

\begin{figure}
\includegraphics[width=0.485\textwidth]{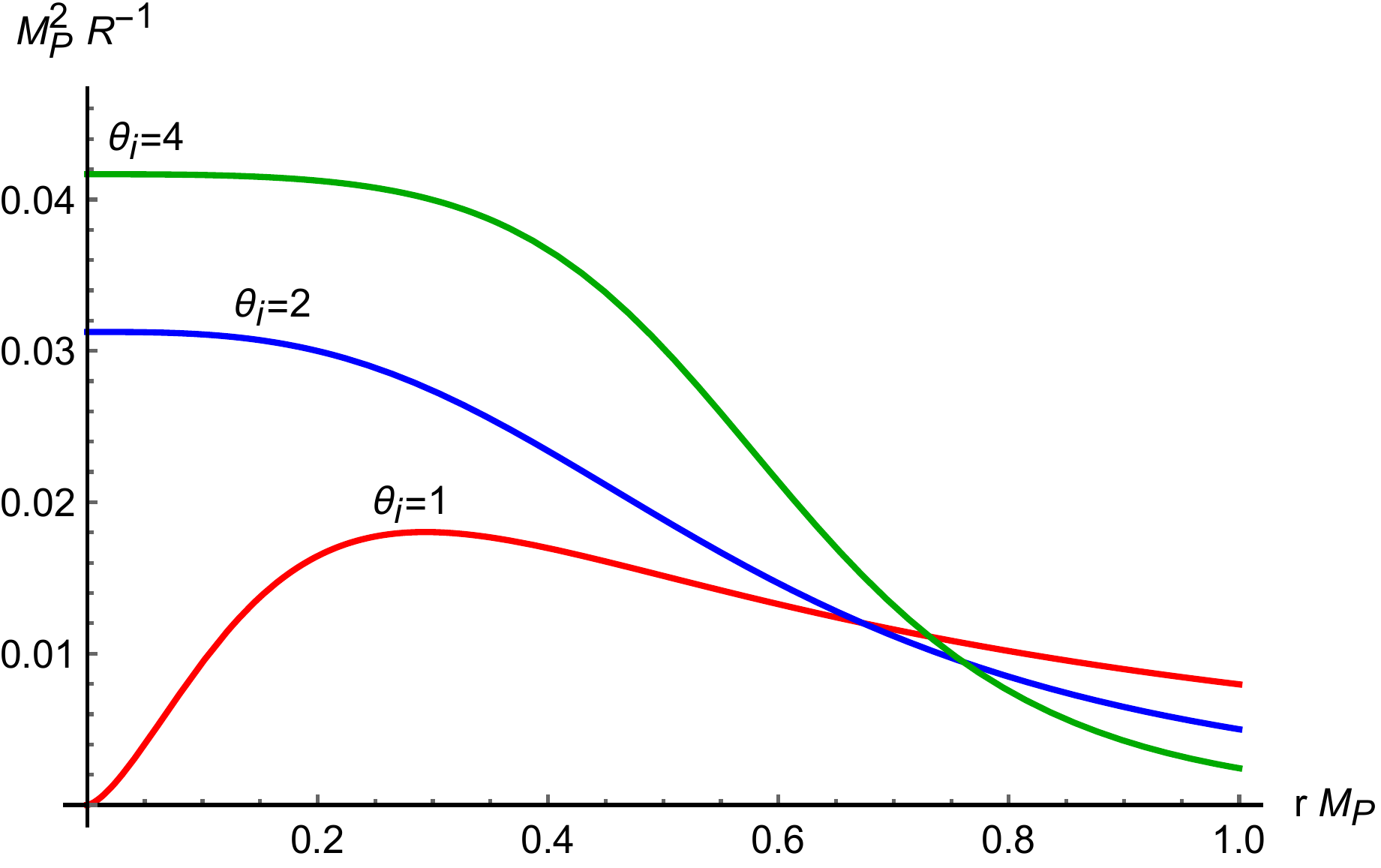}\quad\includegraphics[width=0.49\textwidth]{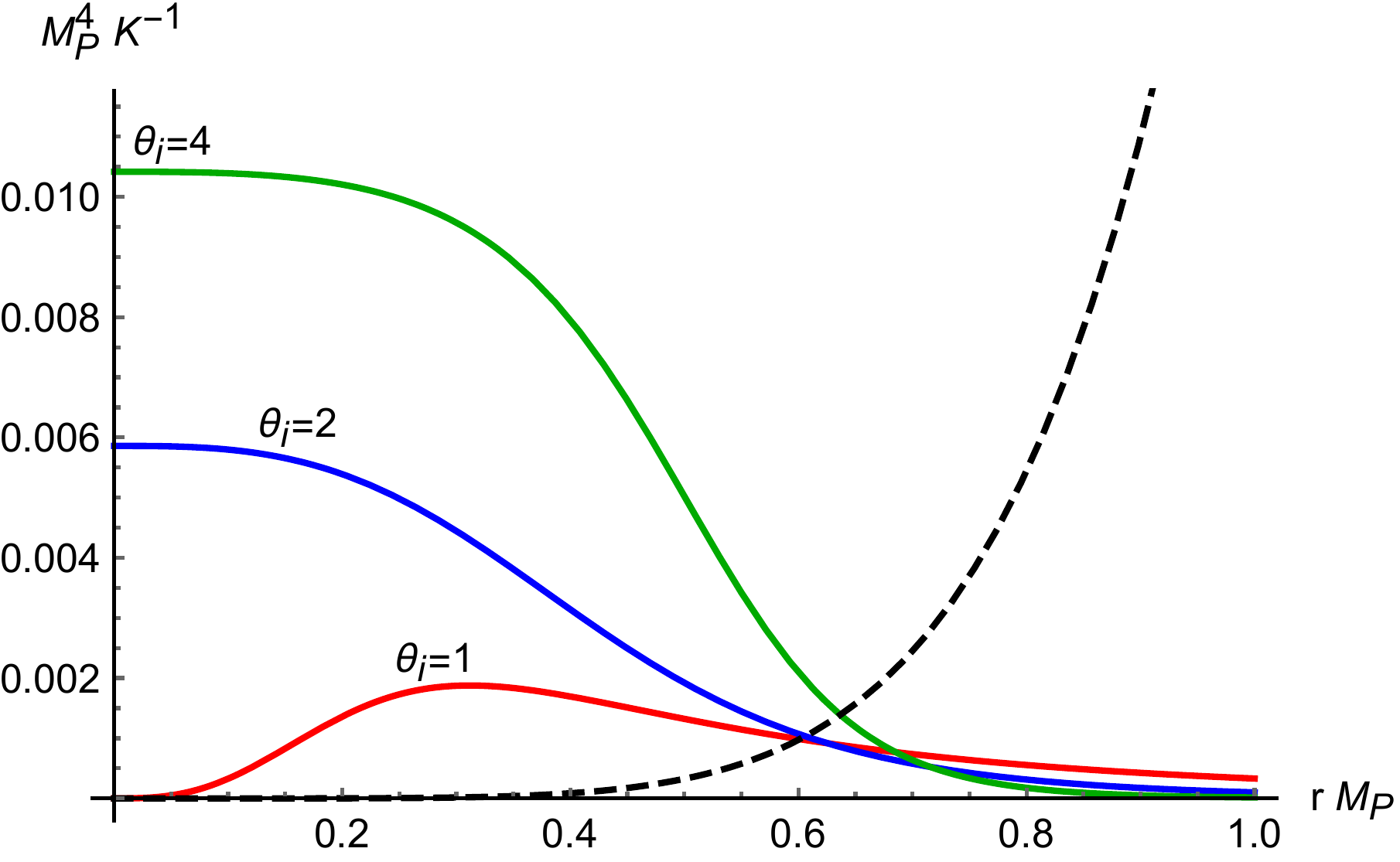}
\caption{Inverse Ricci and Kretschmann scalars for $\lambda_\ast=0$ and different values of the critical exponents $\theta_i$. In the figure we have set $\theta_1=\theta_2$ for simplicity, while all other free parameters are $\sim\mathcal{O}(1)$ in Planck units. The classical Kretschmann scalar (dashed, black line in right panel) is also shown for comparison. The regularity of the curvature invariants crucially depends on the critical scaling \eqref{scalingrc}: only RG flows realizing the conditions $\lambda_\ast=0$ and $\mathrm{Re}[\theta_i]\geq2$ give rise to regular RG-improved spacetimes. \label{fig:scalars}}
\end{figure}

In summary, within the setting explored here, singularity resolution is possible only when the following conditions,
\be \label{constraints1}
\gamma\geq3/2\;\;,\qquad\lambda_\ast=0\;\;,\qquad\theta_1\geq2\;\;,\qquad \theta_2\geq2,
\ee
are satisfied simultaneously (see Fig. \ref{fig:scalars}).
Interestingly, the constraint $\lambda_\ast=0$ is compatible with the condition $\lambda_\ast\leq0$ obtained in the study of RG-improved cosmologies as the key ingredient to realize an inflationary scenario  in agreement with the Planck data \cite{Bonanno:2018gck}.

At this point, two further comments are in order. Firstly, the drastic consequences of a running cosmological constant for the structure of asymptotically safe black holes are not present in unimodular gravity. In this setting $\Lambda$ does not run, as it is a constant of integration that appears at the level of the equations of motion. At the level of the solutions, GR and unimodular gravity are identical. Indications for asymptotic safety in unimodular gravity have been found in \cite{Eichhorn:2013xr,Benedetti:2015zsw,Eichhorn:2015bna}. One can thus explore the potential consequences of unimodular asymptotic safety for singularity resolution by making use of the dynamical equivalence to GR, and then RG-improving the Schwarzschild-deSitter solution with a running $G$ from unimodular asymptotic safety \cite{Torres:2017ygl}. In this case $\Lambda = \rm const$ and $G(k)= k^{-2}(g_{\ast}+ g_1 (k/M_P)^{-\theta_1})$, and an analogous analysis to the case discussed above results in $\gamma \geq 3/2$ and $\theta_1 \geq 0$, in full accordance with the observation that the Newton coupling is relevant at the tentative asymptotically safe fixed point of unimodular gravity.

Secondly, we consider the generalization beyond the Einstein-Hilbert truncation. As discussed before, an extension by $f(R)$ or $R^2, R_{\mu\nu}^2$ does not change the fact that Schwarzschild-(anti)deSitter is a solution to the classical equations of motion and, therefore, only the running Newton coupling and cosmological constant can enter the analysis of the singularity. The inclusion of higher-derivative terms could modify Eq.~\eqref{scalingrc} and \eqref{scalingrc2}, which can feature contributions from further relevant eigendirections of the fixed point. For instance, for the case of real critical exponents, the scaling of the dimensionless cosmological constant becomes
\be
\lambda(k) = \lambda_{\ast} +\sum_i \lambda_i \left( \frac{k}{M_P}\right)^{-\theta_i},
\ee
where the sum extends over all relevant directions for which the corresponding eigenvector has overlap with $\lambda$. It is straightforward to see that the condition for singularity resolution in this case requires the \emph{minimum} of the positive critical exponents to satisfy
\be\label{eq:genconstr}
\theta_{\rm min}\equiv \underset{\mathrm{Re}[\theta_i]>0}{\mathrm{Min}}\big\{\mathrm{Re}[\theta_i]\,\big\}\geq2\;\;.
\ee
This condition generalizes eq.~\eqref{eq:condtheta} to the case of Schwarzschild-deSitter black holes in higher-derivative gravity, and is valid for an arbitrary number of relevant directions. In addition, as already pointed out, the constraint~\eqref{eq:genconstr} does not depend on the particular scale-identification~\eqref{scalk}, and applies to the case of real as well as complex critical exponents. Note that irrelevant critical exponents remain unconstrained in our setting.

\subsection{Case of complex critical exponents}

The approach to the fixed-point regime could also be described by a complex pair of critical exponents. In this case, the modifications of the lapse function are encoded in
\begin{subequations}
\begin{align}
&\phi_G(r)\equiv \frac{m}{\tilde{\xi}^{2}M_{\rm P}^2}{\left(g_\ast+\, \left(g_1 \cos(\theta^{\prime\prime} \log(\tilde{\xi} \, r^{-\gamma})) + g_2 \sin(\theta^{\prime\prime} \log(\tilde{\xi} \, r^{-\gamma})) \right) \tilde{\xi}^{-\theta'} r^{\gamma\,\theta'}\right) \, r^{2\gamma-1}}\;\;, \\
&\phi_\Lambda(r)\equiv \frac{\tilde{\xi}^2 M_{\rm P}^2}{6} \left(\lambda_\ast+\, \left(\lambda_1 \cos(\theta^{\prime\prime} \log(\tilde{\xi} \, r^{-\gamma})) + \lambda_2 \sin(\theta^{\prime\prime} \log(\tilde{\xi} \, r^{-\gamma})) \right) \tilde{\xi}^{-\theta'} r^{\gamma\,\theta'}\right) \, r^{2-2\gamma}\;\;.
\end{align}
\end{subequations}
The latter relations entail a microscopically oscillating Schwarzschild-deSitter spacetime. Employing the  strategy used in the previous paragraph results in a set of constraints similar to Eq.~\eqref{constraints1}
\be \label{constraints2}
\gamma\geq3/2\;\;,\qquad\lambda_\ast=0\;\;,\qquad\theta^\prime\geq2\;\;.
\ee
Independently of the nature of the critical exponents (complex or real), the following conclusion can be drawn:
although the microscopic cosmological constant has to be zero in order to avoid the re-appearance of the classical singularity, nothing prevents it to re-emerge dynamically for $k< \infty$ $(r>0)$. The regularity of the spacetime instead  depends on \textit{how} the running gravitational couplings flow away from the interacting fixed point.
In particular, $\theta_{1,2}\geq 2$ or $\theta'\geq 2$ imply that both the Newton coupling and the cosmological constant are relevant. For the case of the cosmological constant with vanishing fixed-point value, $\theta\geq2$ implies a quadratic or stronger suppression at high scales. One can translate the relevance of $g$ and $\lambda$ into the statement that the two associated scales, namely the transition scale to the fixed-point regime and the mass-scale associated to the cosmological constant, are free parameters. Conversely, this implies that singularity resolution within our setting does not depend on specific values for these two scales, but works for arbitrary choices of these.

\subsection{Singularity-free black holes in a deSitter universe}\label{sec:fullscaleid}

In this section, we spell out how scale-settings $k=k(r)$ involving physically-relevant high-energy scales of the system
lead to $\gamma=3/2$, thereby satisfying the constraint derived previously. For instance, following \cite{br00,2006alfiomartin}, $k$ can be related to the inverse of the classical proper distance
\be\label{ID}
k= \frac{\xi}{d_0(r)}\;\;. 
\ee
The physical scale-setting~\eqref{ID} relies on the fact that the strength of quantum gravitational effects is expected to decrease with the radial geodesic distance from the classical singularity; this implies that the infrared cutoff $k$ should decay as an inverse power of the radial coordinate $r$. Ignoring diffeomorphism invariance,  one would be led to the ``natural'' scale-identification $k\sim r^{-1}$ \cite{br00,2006alfiomartin,Dittrich:1985yb} by dimensional arguments. Although the latter scale-setting works well in the case of quantum theories on a fixed Minkowski background \cite{Dittrich:1985yb}, in the case of GR the radial coordinate $r$ needs to be replaced by its diffeomorphism-invariant counterpart, i.e., the radial geodesic distance $d_0(r)$. In proximity to $r=0$, the classical proper distance scales as $d_0(r)\propto r^{3/2}$ so that
\be \label{kkscaling}
k\simeq \xi\, \sqrt{m\,G_0\,}\, r^{-3/2}\;\;.
\ee
Herein, the prefactor $\sqrt{m\, G_0}$ comes out of $d_0(r)$. In more elaborate settings, where one aims at exploring $m\ll M_P$ (here we focus on the case of non-Planckian black holes, $m\gg M_{\rm P}$), one can choose $\xi = \xi(m/M_{\rm P})$. Then one can accommodate $k \propto m^{-1}$, such that $m/M_{\rm P}\ll1$, as expected, probes the deep quantum gravity regime, $k\rightarrow \infty$.
Note that the scaling relation~\eqref{kkscaling} corresponds to a cutoff function of the form \eqref{scalk} with $\gamma=3/2$ and thereby satisfies the bound in eq.~\eqref{constraints1}.

The strength of the classical gravitational field, encoded in the classical Kretschmann scalar $K_0$, provides another relevant momentum scale of the system \cite{Pawlowski:2018swz}. By dimensional arguments, one can  extract the following alternative scale-identification
\be\label{eq:IDkre}
k=\xi \, K_0^{1/4}\;\;.
\ee
This also follows from the intuition that quantum-gravity effects set in at high curvature scales.
At this point, using the expression~\eqref{eq:ClassKre} for the classical Kretschmann scalar to leading-order in $1/r$, Eq.~\eqref{eq:IDkre} reads
\be \label{krescaling}
k^4\simeq K_0\propto \frac{m \, G_0}{r^6}\;\;.
\ee
The scale-setting \eqref{eq:IDkre} thus gives rise to the same scaling-relation obtained using the proper distance, eq.~\eqref{kkscaling}. 

At this point we discuss the modifications to the classical metric induced by the running gravitational couplings when a scale-setting of the form~\eqref{kkscaling} is employed.
Once the approach to the fixed-point regime, encoded in the critical exponents, is taken into account, the fixed-point lapse function \eqref{treelapse} acquires new contributions. In particular, combining the running couplings in Eq.~\eqref{scalingrc} with the scaling relation~\eqref{kkscaling}, yields the following modified lapse function 
\be
\label{eq:fulllapse}
f(r) = 1-\frac{2mr^2}{\tilde{\xi}^{2}M_\mathrm{P}^{2}} \left(g_\ast+(g_1 \tilde{\xi}^{-\theta_1})\,r^{3\theta_1/2}+(g_2 \tilde{\xi}^{-\theta_2})\,r^{3\theta_2/2}\right)%
-\frac{\tilde{\xi}^{2}M_\mathrm{P}^{2}}{3r} \left(\lambda_\ast+(\lambda_1 \tilde{\xi}^{-\theta_1})\,r^{3\theta_1/2}+(\lambda_2 \tilde{\xi}^{-\theta_2})\,r^{3\theta_2/2}\right).
\ee
Setting $g_{i}=\lambda_{i}=0$ provides the case explored in \cite{ko14} and reported in eq. \eqref{treelapse}. As argued in \cite{ko14}, if $\lambda_\ast \neq 0$, the singularity cannot be removed and it is as strong as in the classical case. However, as is clear from  Eq.~\eqref{eq:fulllapse}, a running cosmological constant $\Lambda(r)$ can be dynamically generated by the flow away from the fixed point. In this way, as long as the condition \eqref{eq:genconstr} on the critical exponents is satisfied, it is possible to reconcile the hints for singularity resolution in RG-improved black holes with the existence of a cosmological constant in the IR limit.

\subsection{Comparison to the literature}\label{sec:comparison}

Instead of plugging in a specific result for an RG trajectory for the running couplings in the Einstein-Hilbert  truncation, we have taken a broader point of view and derived constraints on the behavior of such trajectories in the trans-Planckian regime. Comparing to results in the literature provides a hint whether asymptotically safe quantum gravity might lead to singularity resolution.

We preface the comparison with a cautionary remark, highlighting that our bounds are derived under the assumption that Lorentzian quantum gravity is indeed asymptotically safe, and that RG improving the metric indeed captures quantum gravity effects close to the classical black hole singularity.

In the following, we quote results also from higher-order truncations. As highlighted in Sect.~\ref{sec:realtheta}, the presence of higher-order operators does not prevent Schwarzschild(deSitter) from being a solution to the field equations; in such extensions of the truncation, further relevant directions are produced, and the modifications to the critical scaling of $\lambda$ result in additional constraints on the corresponding critical exponents.

Where results are available that distinguish the fluctuation metric from the background metric, we focus on background couplings only, as these determine the properties of the effective background spacetime. The results for the critical exponents in various truncation schemes computed in the literature are summarized in Tab.~\ref{tab:FPresults}.

\begin{table}[!t]
\centering
\begin{tabular}{|c||c||c||c|c|c|c|}
\hline\hline
Ref. & truncation scheme & specifics & $\;$critical exponents$\;$ & $\theta_1$ & $\theta_2$ & $\theta_3$\\\hline\hline
\cite{Dona:2013qba} & EH & gauge: $\beta=\alpha=1$, cutoff type II& real & $\;\,$3.32$\;\,$ & 1.93 & - \\ \hline
\cite{Gies:2015tca}& EH & $\,$gauge: $\beta=\infty$, field-redefinitions (exp. par.)$\,$ & real & 4 & 2.15 & -\\ \hline
\cite{Gies:2015tca}& EH & gauge: $\beta=\alpha=1$, field-redefinitions & complex & \multicolumn{2}{c|}{$1.69\pm 2.49\,i$} & -\\ \hline
\cite{Gies:2015tca}& EH & gauge: $\beta=\alpha=1$, no field-redefinitions & complex & \multicolumn{2}{c|}{$1.80\pm 2.35\,i$} & -\\ \hline
\cite{Gies:2015tca}& EH & gauge: $\beta=\alpha=0$, no field-redefinitions & complex & \multicolumn{2}{c|}{$\;1.99\pm 3.06\,i\;$} & - \\ \hline
\cite{Gies:2015tca}& EH & gauge: $\beta=1,\,\alpha=0$, field-redefinitions & complex & \multicolumn{2}{c|}{$2.03\pm 2.69\,i$} & -\\ \hline
\cite{Gies:2015tca}& EH & gauge: $\beta=\infty, \,\alpha=0$, field-redefinitions & complex & \multicolumn{2}{c|}{$2.25\pm 2.79\,i$} & -\\ \hline
\cite{Falls:2014tra} & $f(R)$ to $R^{34}$& gauge: $\beta=\alpha=0$ & complex & \multicolumn{2}{c|}{$2.51\pm 2.41\,i$} & 1.61\\ \hline
\cite{Benedetti:2009rx} & $\mathbf{1}$, $R$, $R^2$, $R_{\mu\nu}R^{\mu\nu}$ & gauge: $\beta=\alpha=0$ & real & 8.40 & 2.51 & $\;$1.69$\;$\\ \hline\hline
\cite{Dona:2013qba} & EH with SM matter & gauge: $\alpha=\beta=1$, type II & real & 3.92 & 1.65 & - \\ \hline
\cite{Dona:2013qba} & EH with SM matter & gauge: $\alpha=\beta=1$, type Ia & real & 3.92 & 2.19 & - \\ \hline
\cite{Alkofer:2018fxj} & $\;$$f(R)$ to $R^9$ with SM matter$\;$ & gauge $\alpha=0, \beta = -\infty$, type I (exp. par.) & real & 4 & 2.28 & - \\ \hline\hline
\end{tabular}
\caption{\label{tab:FPresults} We show the results for the critical exponents in various studies of the gravitational RG flow of the background couplings. Note that only critical exponents $\theta_i$ associated to relevant directions are shown.  The information provided in the column ``specifics'' refers to details of the RG setup which are explained in the corresponding references. For our purpose it is sufficient to note that the dependence of the universal critical exponents on these unphysical parameters is caused by the projection of the flow on finite-dimensional subspaces of the theory space (column ``truncation''), and provides a rough measure of the systematic error arising in the corresponding truncation scheme.
}
\end{table}

Note that the variation of the universal critical exponents with changes in the regulator and truncation scheme provides a rough estimate of the systematic error in the calculation of these quantities. Within this rough estimate,  we tentatively conclude that the critical exponents $\theta_{1,2}$ and $\theta^\prime$ could be compatible with the conditions~\eqref{constraints1} and~\eqref{constraints2}. However, as the critical exponents derived in the literature are such that, roughly, $\mathrm{Re}[\theta_\mathrm{min}]\simeq 2\pm0.5$, in order to establish whether or not asymptotically safe quantum gravity satisfies our constraints, computations of critical exponents from extended truncations are required. In addition, a vanishing fixed-point value for the cosmological constant has not been encountered in the literature so far.

\section{Conclusions} \label{sect4}
 Understanding the structure of black holes in quantum gravity constitutes a fundamental advancement towards testing quantum-gravity models. The first step in this direction is a consistency check, namely that quantum-gravity effects resolve or suitably weaken the classical singularity. This provides a starting point for a well-defined description of black holes in our universe and, most importantly, opens the door towards potential future observational tests, where gravitational-wave data on black-hole mergers could be exploited to constrain quantum gravity. Specifically, quantum gravity might not only modify the spacetime close to the classical singularity, but also leave further imprints in a black hole's structure, e.g., in its quasi-normal mode spectrum,  see, e.g., \cite{Cai:2015fia,Kokkotas:2017zwt} for the case of quadratic gravity. In effect, quantum-gravity contributions generate additional terms in the effective action. Therefore, all tests of modifications of gravity implicitly constrain quantum-gravity models.

Here, we undertake one step towards reaching the first goal, and test whether an asymptotically safe regime could
lead to singularity resolution.  Scale-invariance implies that the Newton coupling falls off quadratically with the momentum scale beyond the transition scale to the fixed-point regime. Thus one might expect at least a weakening of the classical singularity, associated with the weakening of gravity due to scale invariance. Using an RG improvement procedure, we show how to upgrade the classical Schwarzschild-deSitter solution to a modified, singularity-free spacetime. Specifically, this is obtained by promoting the Newton coupling and cosmological constant in the classical solution to their running counterparts and then identifying the RG scale $k$ with a physical scale of the system.
Although the RG improvement procedure does not provide a strict derivation of the quantum metric from asymptotically safe gravity, the large symmetry of Schwarzschild-deSitter spacetimes effectively provides a single high-energy scale that is suitable for an identification of the  cutoff function $k(r)$ in the UV. Specifically, this energy scale is given by the Kretschmann scalar, $K= R_{\mu\nu\kappa\lambda}R^{\mu\nu\kappa\lambda}$, and can be equivalently traded for the radial geodesic distance.

The requirement of singularity-resolution results in three conditions on the gravitational RG flow. The first one concerns the scale dependence of the cutoff function $k(r)$, and it is satisfied by the most straightforward choice of scale identification based on physical arguments and dimensional analysis, namely $k^4 \sim K_0$  or, equivalently, $k\sim d_0^{-1}(r)$. The second condition requires the microscopic fixed-point value of the cosmological constant to vanish \cite{ko14}. The observed low-energy value of the cosmological constant can then be generated by the RG flow away from the scale invariant regime.
To account for this, in our study we focused not only on the fixed-point regime but also on the linearized flow of couplings in the proximity of the fixed point.
This results in a third constraint, which is a condition on the critical exponents that characterize the approach to the fixed-point regime.
Our results imply that the real part of all relevant
critical exponents, whose eigendirection has overlap with the cosmological constant, should satisfy the bound $\theta_i\geq 2$. Combining the latter constraint  with the conditions to obtain an RG-improved inflationary cosmology compatible with the Planck data \cite{Bonanno:2018gck}, provides a very strict restriction on the value of the least (relevant) critical exponent, namely, $2\leq\theta_\mathrm{min}\leq4$.
Intriguingly, results on the critical exponents in studies of the truncated, Riemannian RG flow provide values close to this bound. Taking the variation of the universal critical exponents under changes of truncation, gauge and regulator function as a rough estimate of a systematic error, we conclude that the critical exponents obtained in the literature so far could be compatible with our constraint, but future studies based on extended truncations are required to establish whether this bound is indeed fulfilled. Our work paves the way towards understanding the structure of regular black holes in asymptotically safe quantum gravity in a setup that is observationally viable in that it accounts for a nonzero infrared value of the cosmological constant.

In the absence of a running cosmological constant (e.g., in unimodular gravity), the condition for singularity resolution becomes $\theta>0$ for the critical exponent of the Newton coupling. This condition ensures that an ultraviolet fixed point is reached by the running Newton coupling. In contrast, it is violated in perturbative quantum gravity, where $\theta=-2$ is due to the canonical scaling at the free fixed point. In a quantum field theoretic setting for gravity, singularity resolution thus appears to be tied to non-perturbative renormalizability, also known as asymptotic safety. There, the antiscreening character of quantum-gravity fluctuations could lead to a sufficient weakening of the gravitational interaction at high curvature scales and might result in regular black holes.

\begin{acknowledgments}
We thank Alfio Bonanno, Kevin Falls, Jan M.~Pawlowski and Dennis Stock for discussions.
This research is supported by the DFG under grant no Ei-1037/1.
\end{acknowledgments}

\bibliography{fr}

\end{document}

%% file: jdefs
\def\apjl{ApJ}%
\def\apjs{ApJS}%
\def\ao{Appl.\ Opt.}%
\def\aap{A\&A}%
\def\jcap{JCAP}%
\def\prd{Phys.\ Rev.\ D}%